\documentclass[aps,prl,reprint,superscriptaddress,longbibliography]{revtex4-1}
\usepackage{graphicx}
\usepackage{graphics}
\usepackage{epsfig}
\usepackage{epstopdf}
\usepackage{placeins}
\usepackage{float}
\usepackage{array} 
\usepackage{multirow}
\usepackage[english]{babel}

\usepackage{xcolor}
\usepackage{amsmath}

\newcommand{\DTG}{$\mathrm{Dy_2GdN@C_{80}}$}
\newcommand{\DTS}{$\mathrm{Dy_2ScN@C_{80}}$}
\newcommand{\DDL}{$\mathrm{Dy_2LuN@C_{80}}$}
\newcommand{\DDD}{$\mathrm{Dy_3N@C_{80}}$}
\newcommand{\DSS}{$\mathrm{DySc_2N@C_{80}}$}
\newcommand{\DDS}{$\mathrm{Dy_2ScN@C_{80}}$}

\begin{document}

\title{Gadolinium as a single atom catalyst in a single molecule magnet}

\author{Aram Kostanyan}
\affiliation{Physik-Institut, Universit\"at Z\"urich, Winterthurerstrasse 190, CH-8057 Z\"urich, Switzerland}
\affiliation{Swiss Light Source, Paul Scherrer Institut, CH-5232 Villigen PSI, Switzerland}
\author{Christin Schlesier}
\affiliation{Leibniz Institute of Solid State and Materials Research, Dresden, D-01069 Dresden, Germany}
\author{Rasmus Westerstr\"om}
\affiliation{Division of Synchrotron Radiation Research, Institute of Physics, SE-221 00 Lund, Sweden}
\author{Jan Dreiser}
\affiliation{Swiss Light Source, Paul Scherrer Institut, CH-5232 Villigen PSI, Switzerland}
\author{Fabian Fritz}
\affiliation{Department of Physics, University Osnabr\"uck, D-49076 Osnabr\"uck, Germany}

\author{Bernd B\"uchner}
\affiliation{Leibniz Institute of Solid State and Materials Research, Dresden, D-01069 Dresden, Germany}
\author{Alexey A. Popov}
\affiliation{Leibniz Institute of Solid State and Materials Research, Dresden, D-01069 Dresden, Germany}
\author{Cinthia Piamonteze}
\affiliation{Swiss Light Source, Paul Scherrer Institut, CH-5232 Villigen PSI, Switzerland}
\author{Thomas Greber}
\email{greber@physik.uzh.ch}
\affiliation{Physik-Institut, Universit\"at Z\"urich, Winterthurerstrasse 190, CH-8057 Z\"urich, Switzerland}


\date{\today}

\begin{abstract}
Endohedral fullerenes are perfect nanolaboratories for the study of magnetism.
The substitution of a diamagnetic scandium atom in \DTS\ with gadolinium decreases the stability of a given magnetization and demonstrates Gd to act as a single atom catalyst that accelerates the reaching of thermal equilibrium.
X-ray magnetic circular dichroism at the M$_{4,5}$ edges of Gd and Dy shows that the Gd magnetic moment follows the sum of the external and the dipolar magnetic field of the two Dy ions and compared to \DTS\ a lower exchange barrier is found between the ferromagnetic and the antiferromagnetic Dy configuration. 
The Arrhenius equilibration barrier as obtained from superconducting quantum interference device magnetometry is more than  one order of magnitude larger, 
though a much smaller prefactor imposes faster equilibration in \DTG.
This sheds light on the importance of the angular momentum balance in magnetic relaxation.                                                                                                                                                                              

\end{abstract}

\maketitle

A catalyst accelerates the approach of thermal equilibrium without being consumed by this process.
In the classical picture it lowers the kinetic barrier between two states A and B, where the transition rates are described by Boltzmann factors comprising the barrier, the energy difference between A and B, and by a prefactor reminiscent to the attempt frequency.
The concept of a catalyst may as well be applied for the case of single molecule magnets that were prepared in a state outside thermal equilibrium towards which they decay with certain rates.
From the temperature dependence of these rates Arrhenius barriers and attempt frequencies may be inferred.
The exponential prefactors in the kinetics reflect the energy landscape where a process runs. 
In the present case the 4f electron spins
of a single gadolinium atom are shown to catalyze magnetic transition rates dramatically.                     

Single molecule magnets (SMM's) display hysteresis, i.e. maintain non-equilibrium magnetization for a measurably long period of time \cite{Sessoli1993, Gatteschi2006, cor2019}.
After the discovery of single ion molecule magnets \cite{ish2003} the lanthanide magnetochemistry got a significant boost, where new strategies for the improvement were developed \cite{woo2013,lid2015,liu2018}. 
With these ideas on rational design magnetic hysteresis of single molecules at 60~K \cite{goo2017} and even above liquid nitrogen temperatures \cite{guo2018} is reached by now.
Endohedral fullerenes provide a versatile environment to encapsulate different diamagnetic and paramagnetic ions \cite{ste99,Popov2013}.
First magnetic studies were performed on Gd@C$_{82}$ \cite{fun1995} and it took 17 more years until SMM behaviour in endohedral fullerenes was found with \DSS \cite{Westerstrom2012}. 
In the following many different C$_{80}$ mixed dysprosium-lanthanide nitride clusters have been synthesized and magnetically characterized \cite{Junghans2015,Schlesier2019,Spree2020}.
As in other radical bridged lanthanide complexes \cite{rin2011} dysprosium ion pairs appear to form excellent SMM's \cite{spr2019}.
For the case of \DTS\ the particularly long zero field lifetime of the magnetization was attributed to exchange protection of two Kramers ions \cite{Westerstrom2014}.
Heterometallic clusterfullerenes require more effort for synthesis and separation, and only recently an endeavour of implementing three different rare earth atoms inside C$_{80}$ combined single atom magnetism and luminescence \cite{Nie2019}.
If more than one paramagnetic atomic species is involved, element specific methods such as x-ray magnetic circular dichroism (XMCD) \cite{van2014} give unprecedented insight into the magnetic ordering of heteroatomic clusters \cite{dre2012,cor2012,luz2013}.

Here we investigate the interplay between two Dy and one Gd ion in a 1 nm carbon cage. 
The substitution of scandium with gadolinium
decreases the excitation gap between the ferro- and antiferromagnetically coupled Dy doublets. 
Although the kinetic barrier {\it{increases}}, the prefactor  dictates a lower magnetization lifetime in \DTG . 
This strong change in magnetization dynamics indicates Gd to break the spin flip protection symmetries which are found in \DTS\ and exhibits the role of the Gd angular momentum in the change of magnetisation. 
      
      \DTG\ endofullerenes were produced by arc-discharge synthesis using the corresponding metals \cite{ste99,Popov2013}. 
The subsequent separation by high pressure liquid chromatography resulted in a sample of 95 \% purity with 5\% \DDD\ contamination as inferred from time of flight mass-spectrometry \cite{supplementals}.
The \DDD\  content does not affect the conclusions of this paper.
The magnetization measurements were performed by x-ray magnetic circular dichroism (XMCD) and SQUID magnetometry.
XMCD was performed at the X-Treme beamline of the Swiss light source (SLS) \cite{Piamonteze2012} 
on a sample obtained after spray coating of a toluene solution of the molecules on an aluminum substrate.
The total electron yield 
was normalized by the secondary electron current from a gold mesh in the x-ray beam before it hits the sample and by the x-ray absorption cross section of gold.
The SQUID magnetometer was a Quantum Design (QD)  MPMS3 Vibrating Sample Magnetometer (VSM), where we performed magnetization experiments in external magnetic fields up to 7~T and temperatures down to 1.55~K. 
For the SQUID measurements, the toluene solution was drop-cast into a QD polypropylene powder sample holder. 
A temperature independent diamagnetic background of $-5.8\times 10^{-8}$~Am$^{2}$/T was inferred from temperature dependent magnetization measurements.
From the paramagnetic saturation at 2 K and 7 T, m$_{sat}$ of \(2.17 \times 10^{-6}\)~Am$^{2}$ and a maximum magnetic moment of 17 $\mu_B$ an ensemble of \(1.4 \times 10^{16}\) molecules with a mass of 33~$\mu$g is inferred. (The saturation magnetic moment of 17 $\mu_B$ being the sum of  two non-collinear Dy $J_z$=15/2 and one collinear Gd $J_z$=7/2 moment.)
                                                     
\begin{figure}[ht]
    \centering
    \includegraphics[width=0.8\columnwidth]{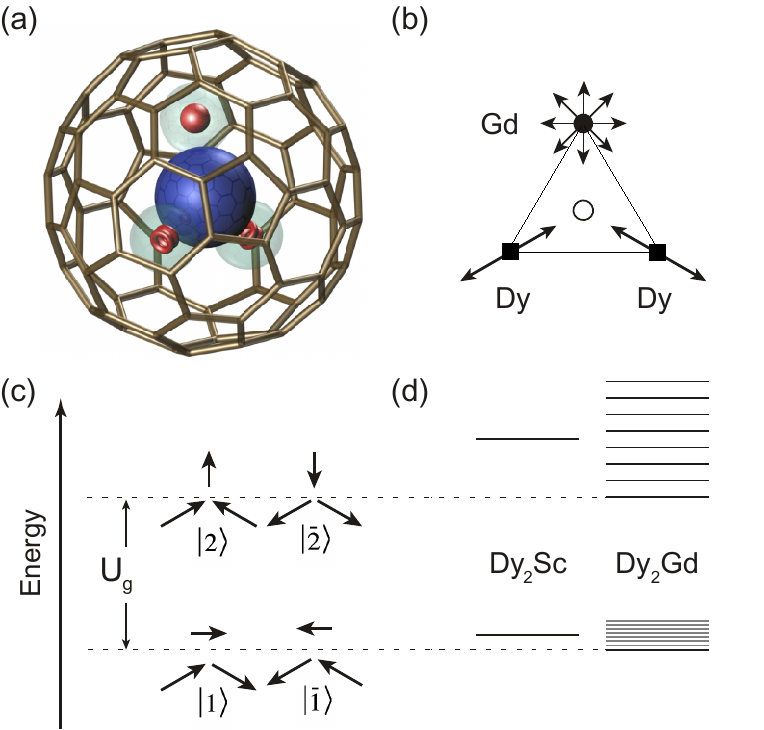}
    \caption{(a) Ball and stick model of \DTG\ with a van der Waals diameter of about 1.1 nm. 
    The endohedral ions are mimicked with their ion radii. In the center of the lanthanides (turquois) the 4f $J_z=15/2$ and $J_z=7/2$ orbitals for Dy and Gd are depicted in red. 
    (b) Sketch of the endohedral unit with the orientations of the magnetic moments. The two-way arrows indicate the possible orientations of the Dy\(^{3+}\) moments that are aligned along the N--Dy axes, while the Gd moment is oriented along the total magnetic field.
    (c) Ground state magnetic structure at zero external magnetic field, as an extension of the pseudospin model for \DTS\ \cite{Westerstrom2014}. The orientation of the Gd\(^{3+}\) moment is assumed by the B-field of the two Dy moments. $U_g$ is the energy difference between the lowest ferromagnetic ($|1\rangle$, $|{\bar{1}}\rangle$) and the lowest antiferromagnetic states ($|2\rangle$, $|{\bar{2}}\rangle$).  
    (d) Energy levels of \DTS\ and \DTG. The 8 electron spin states of Gd are split by the B-field of the Dy ions. Note: Here the average ferro- antiferro- splitting is assumed to be equal for both molecules, which turns out not to be the case.}
\label{F1}
\end{figure}

Figure~\ref{F1} shows the ground state model for \DTG\ at zero external magnetic field. 
The electric field of the nitrogen ion  lifts the 16-fold degeneracy of the Dy $^{6}$H$_{15/2}$ Hund ground state \cite{Jiang2011, Westerstrom2014}, while no ligand field splitting is expected for the half filled 4f shell of the Gd $^{8}$S$_{7/2}$ configuration.
Since the ligand field interaction is much stronger than mutual magnetic interaction, it is expected that the Dy ions assume like in \DDS~ $J_z \pm 15/2$ states.
At low temperatures the higher lying states in the $J_z$ manifold of dysprosium may be neglected and the picture of two pseudospins applies \cite{Westerstrom2014}.
In zero field the two pseudospins of the Dy ions arrange in two time reversal symmetric doublets $|1\rangle$, $|{\bar{1}}\rangle$ (ferromagnetc), and $|2\rangle$, $|{\bar{2}}\rangle$ (antiferromagnetic). 
The two Dy atoms in the ferromagnetic configuration create at the Gd site a dipole-field of 180~mT parallel to the Dy--Dy axis, while in the antiferromagnetic configuration it is $720$ mT and directed perpendicular to the Dy--Dy axis.
\begin{figure}[]
    \centering
    \includegraphics[width=0.9\columnwidth]{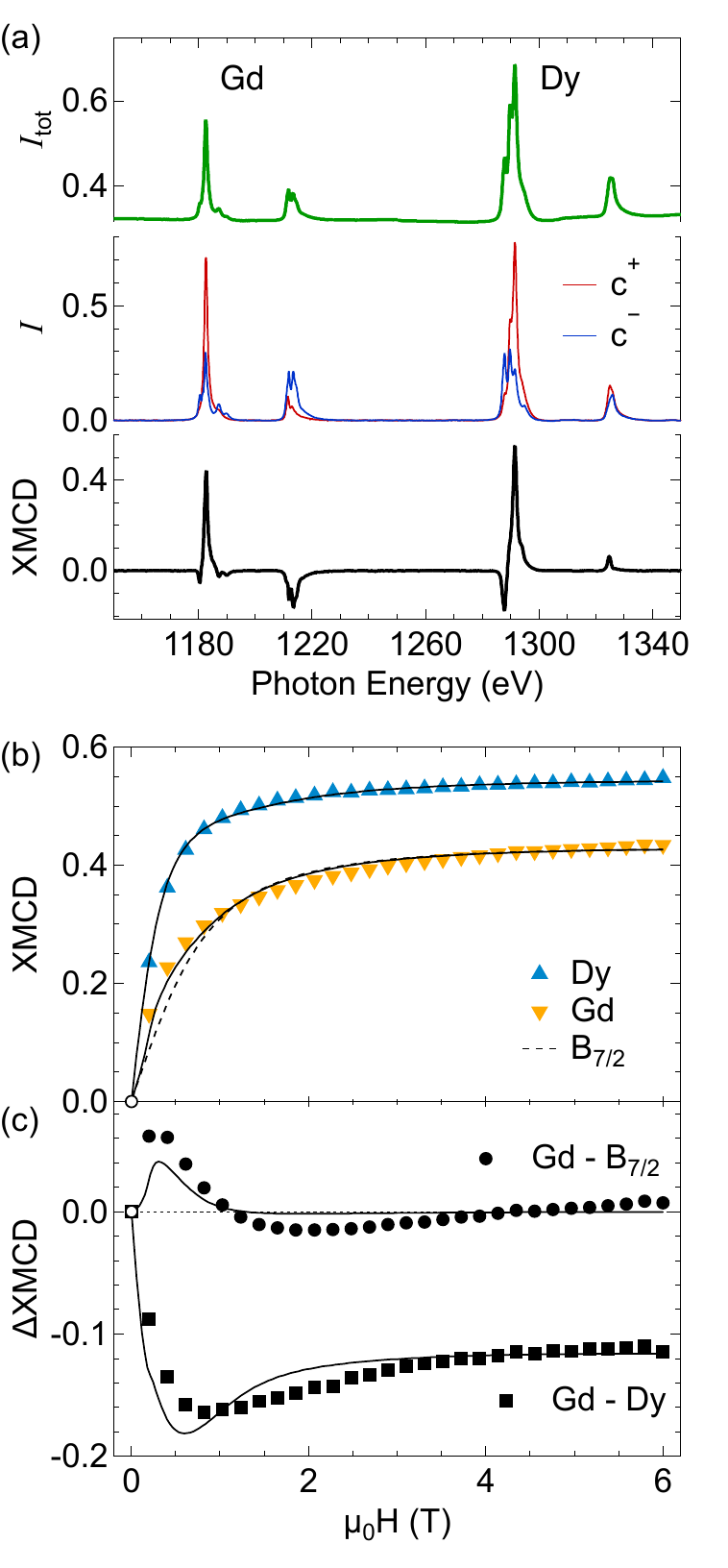}

    \caption{(a) Top: X-ray absorption spectrum $I_{\mathrm{tot}}$ vs. photon energy of \DTG\ (green line). The intensity ratio between Gd and Dy is in line with the 1:2 stoichiometry. Maximum electron yield 106~pA.
Middle: Background subtracted x-ray absorption spectra $I(c^{+})$ and  $I(c^{-})$ for both x-ray helicities with the maxima of the Gd and Dy intensities normalized to 1. External field 6.5~T, parallel to the x-ray incidence.
Bottom: X-ray magnetic circular dichroism (XMCD) $I(c^{+})-I(c^{-})$.
(b) Field dependent XMCD of Dy (up-triangle, light blue) and Gd (down-triangle, yellow). Brillouin function $B_{7/2}$ (dashed line) corresponding to $g=2,~J=7/2$, $T=1.95~K$, representing the magnetization of free Gd. The solid lines are fits of the element specific magnetisation curves to an extended pseudospin model \cite{supplementals}.
        (c) Differences between XMCD(Gd) and $B_{7/2}$ (black disks), and XMCD(Dy) (black squares). The solid lines are obtained from the fit results in (b). The peak of XMCD(Gd) - $B_{7/2}$ indicates the B-field of the Dy ions at the Gd site.
}
    \label{F2}
\end{figure}

The ground state as proposed in Figure~\ref{F1} is confirmed by x-ray magnetic circular dichroism.
The XMCD for Dy and Gd is shown in the bottom of Figure~\ref{F1}(a). 
Sum rule analysis \cite{Teramura1996,Thole1992,Carra1993} 
reveals effective saturation magnetic moments of Gd$^{3+}$ and Dy$^{3+}$ in \DTG\ \cite{supplementals}. 
The Dy effective magnetic moment 
at saturation of \mbox{4.5$\pm$0.1 $\mu_B$} compares well with the value in \DSS\ \cite{Westerstrom2012}. The moment of Gd gets 5.8$\pm$0.1 $\mu_B$. It is 
smaller than the expectation for a free, collinear Gd ion as also observed in other Gd molecular magnet systems  
\cite{den04}.

The magnetization curves of Gd and Dy are shown in Figure~\ref{F2}(b). For endohedral Gd it deviates from a Brillouin function with $g=2$ and $J=7/2$.
This certifies endohedral Gd not to behave as a free ion but that it is subject to magnetic interaction with the two Dy ions.
Compared to the free ion, the relative magnetization of Gd in \DTG\ is largest at a field of about 0.2 T, which is close to the field imposed by the two Dy ions in the ferro ground states.

The above statements are substantiated with the implementation of a pseudospin model to the level scheme in Figure~~\ref{F1}(d) \cite{supplementals}.
The solid lines in Figure~~\ref{F2}(b) are the best fit, where the access to the individual magnetisation curves of Dy and Gd improves the reliability of the extracted parameters significantly. 
The magnetic moment of Dy gets 8.8~$\mu_B$ and is in line with that of \DTS ~\cite{Westerstrom2014}, while that of Gd gets 6.9~$\mu_B$. 
The parameter $B_D$ \cite{supplementals} that describes the splitting of the Gd states due to the Dy dipolar fields is 0.24~T, which is close to the value of 0.2~T of a moment of 10~$\mu_B$ at a Dy-Gd distance of 0.36~nm. 
This affirms that in zero field the Gd magnetism is governed by the dipolar fields of the two Dy ions and that possible exchange interaction between Dy and Gd must be much smaller. 
Finally, the excitation energy or gap between the lowest ferromagnetically and the lowest antiferromagnetically coupled states $U_g$ is determined from comparison of the data to the model. In \DTG\ it decreases, compared to \DTS\ from 9 to 0.1$\pm$0.8~$k_B$~K. 
This $U_g$ value points to zero field degeneracy of the four possible Dy spin configurations. It is only partly due to the higher Dy-Gd coupling in the antiferromagnetic states, which accounts for 2.5 $k_B$~K, but also due to a smaller Dy-Dy exchange for the case of \DTG .
Such variations in exchange coupling are difficult to predict but were even observed between \DTS\ and \DDL\, where both, scandium and lutetium are diamagnetic \cite{Spree2020}. 

The smaller barrier $U_g$ between the ferro and the antiferro states is a hint that Gd may accelerate the approach of the Dy spin system to thermal equilibrium.
\begin{figure}[t]
    \centering
\includegraphics[width=0.9\columnwidth]{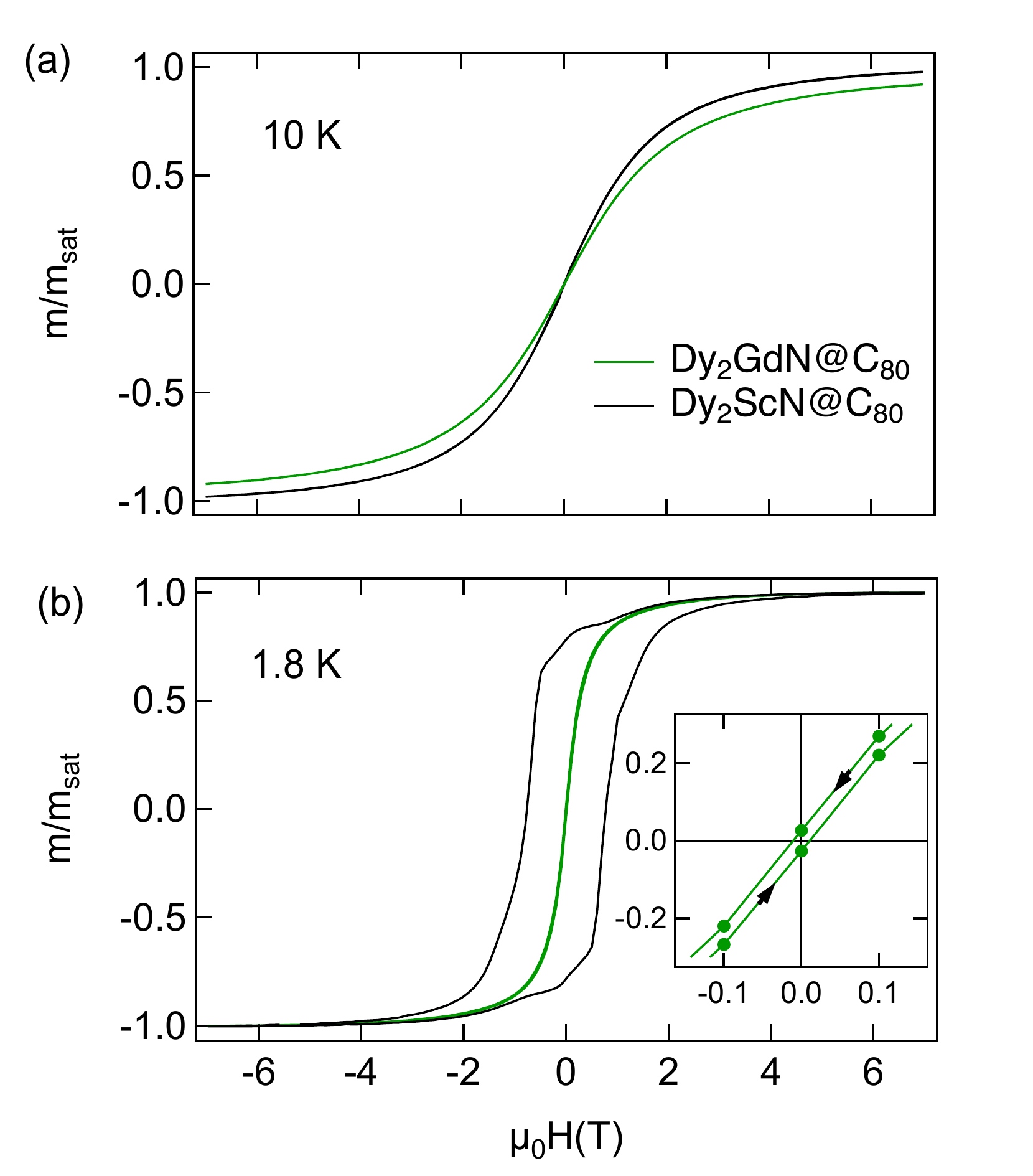}
\caption{Comparison of the magnetization of \DTG\ (green) and \DTS\ (black) normalized with the corresponding saturation values $\rm{m_{sat}}={\rm{m(7~T,2~K}})$. (a) At 10~K both molecules show paramagnetic behaviour, where the initial slope of the normalized magnetization of \DTG\  is a factor of 0.8 smaller than that of  \DTS . 
(b) At 1.8 K  with a field scan rate of 5.3 mT/s \DTG\ shows a small hysteresis of 68~mT~$\rm{m_{sat}}$, while \DTS\ displays a large hysteresis with an area of 3190~mT~$\rm{m_{sat}}$. The inset zooms  the hysteresis of \DTG\ with the axes $\rm{m/m_{sat}}$ and $\mu_0\rm{H(T)}$.  Black arrows indicate the field scan directions.}
\label{F3}
\end{figure}
In Figure~~\ref{F3} magnetization curves for the two molecules are displayed.
At 10~K temperature both display paramagnetism with a characteristic thermal equilibrium curve.
The relative zero field susceptibility 
of  \DTG\ is a factor of 0.81 smaller than for  \DTS . This confirms that more magnetic states are available in \DTG. The additional states of the Gd ion interfere with the Dy$_2$ units and decrease the hysteresis of \DTS\ at 1.8~K by a factor of 47 (see Figure~\ref{F3}(b) and supplementary materials \cite{supplementals}). 
Apparently, the exchange protection as it is operational in \DTS\ breaks down if gadolinium sits in the same cage instead of scandium. This identifies Gd to soften the hysteresis, and makes it an option in engineering of single molecule magnets, if e.g. heat dissipation shall be minimized in high frequency applications.

For all molecules investigated so far,  the ground state parameter $U_g$ of the molecules as determined from the magnetisation in thermal equilibrium was in line with the Arrhenius barrier $\Delta_{\rm{eff}}$ from the decay of the magnetisation \cite{Westerstrom2014,Kostanyan2020}. In the following we will see that this is not the case for \DTG, and the discrepancy between $U_g$ and $\Delta_{\rm{eff}}$ spots new light on the demagnetisation dynamics.
\begin{figure}[h!t]
\begin{center}
\includegraphics[width=\columnwidth]{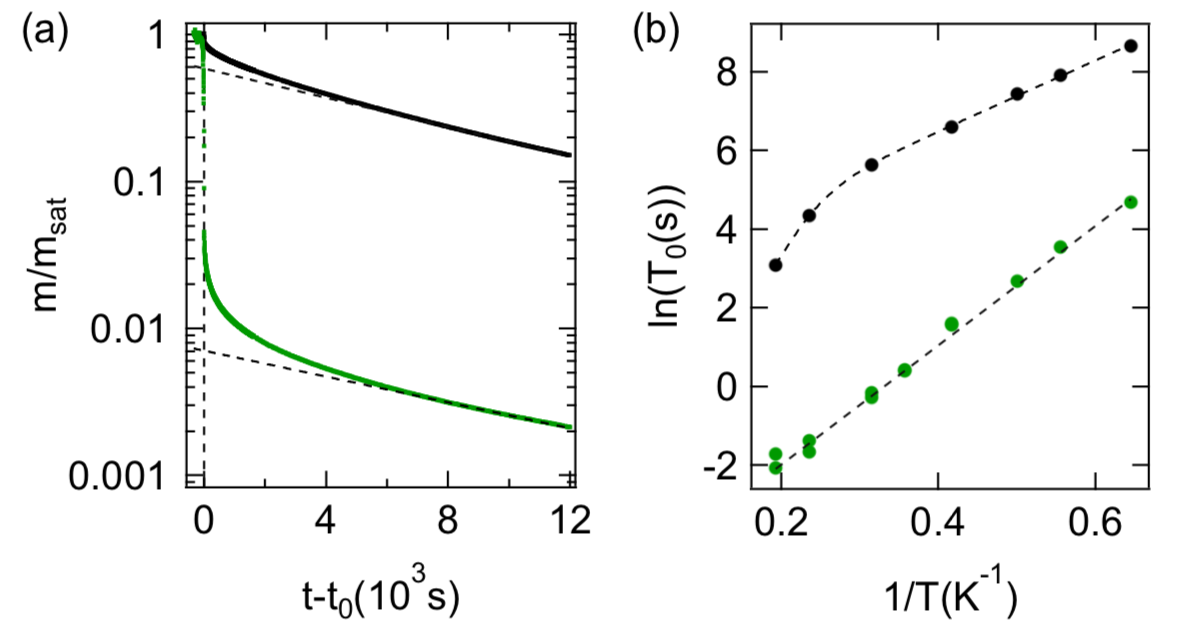}
\caption{Magnetization decay of \DTS\ (black) and \DTG\ (green) after saturating the samples at 7 T and ramping the magnet down to zero field in 350~s. (a) Magnetization as a function of time at 1.55 K. $t_0$ is the time when zero field is reached.  (b) Arrhenius plot of the remanence times $\ln(T_0)$ for the two molecules at temperatures between 1.55 and 5~K. For \DTS\ (black) a low temperature barrier of 9.0$\pm$0.2~K is found \cite{Westerstrom2014}. For \DTG\ (green) a straight line leads 15.1$\pm$0.4~K, with a 1/2500 times smaller prefactor.}
\label{F4}.
\end{center}
\end{figure}
      
The fluctuation rates of the magnetic states are expressed in the kinetics that describe the approach to thermal equilibrium. Figure~~\ref{F4}(a) shows an experiment where the magnetic moments of the molecules were saturated in a magnetic field, 
which is then ramped down to zero. The subsequent decay 
contains information on the ground state and the dynamics of the spin flips involved.
 The decay rate is not constant but decreases to a constant value. This behaviour is common to single molecule magnets \cite{rin2011,Westerstrom2012}. 
At 1.55~K and after two hours the decay rates of \DTS\ and \DTG\ are in the order of 10$^{-4}$s$^{-1}$, though the back-extrapolation of these rates to $t=t_0$ indicates that 60\% of all \DTS\ decayed with this rate, while it were 0.7 \% only in the case of  \DTG . 
For a quantification of the kinetics that describe the approach to thermal equilibrium we investigate the decay rates as a function of  temperature.
It is, however, difficult to determine decay rates below 10$^{-3} \rm{m_{sat}}$, because of the relaxation of the magnet in the SQUID. In order to get a reliable value for the magnetic lifetime we evaluated the integral of the magnetization after reaching zero field at time $t_0$ that we like to call remanence time $T_0$:

\begin{equation}
T_0=\int_{t_0}^\infty {\rm{(m/m_{sat})} }\ dt,   
\end{equation}
where $T_0$ corresponds to the decay time $\tau$ of a single exponential \mbox{$\rm{m}(t-t_0)=\rm{m_{sat}}\exp(-(t-t_0)/\tau)$}. The validity of using this remanence time to describe the decay of the magnetization is justified by the evaluation of the decay data of \DTS (see Figure~\ref{F4}(b)), where a barrier $\Delta_{\rm{eff}}/k_B$ of 9.0$\pm$0.2~K is found as compared to 8.5$\pm0.5$~K
with the standard method \cite{Westerstrom2014,supplementals}. 
This barrier fits the dipole and exchange splitting in \DTS\ as inferred from the equilibrium magnetization  \cite{Westerstrom2014}.
The Arrhenius slopes in Figure~~\ref{F4}(b) indicate for \DTG\  a larger barrier than for \DTS. 
The barrier $\Delta_{\rm{eff}}$ of 15.1$\pm$0.4~K can not be understood with a decay path, where the lowest ferro-states are directly excited into the lowest antiferro-states, since  
$U_g$ is much lower. 
This suggests that the flip of the magnetization in a $|1\rangle\rightarrow|\bar{1}\rangle$ transition involves excitation of the Gd 4f states. The high barrier is consistent with a picture where  the Dy angular momentum change in a $|1\rangle \rightarrow |\bar{1}\rangle$ transition can be temporarily stored in the Gd ions. 
Furthermore, the kinetics of the magnetization decay comprises a 
prefactor $\tau_0$ in
$\tau(T)=\tau_0\exp{(\Delta_{\rm{eff}}/k_BT)}$. 
It is a factor of 4.1$\pm$0.8$\times 10^{-4}$ smaller in \DTG\ as compared to \DTS . 
It includes, besides an attempt frequency, as well the quantum symmetry of the object \cite{Kostanyan2020}.
Apparently, single Gd atoms break the symmetry that imposes the exchange protection in \DTS\  and act as a reservoir for angular momentum that is involved in the flip of the magnetization.
 
In conclusion x-ray magnetic circular dichroism establishes non-collinear magnetism of the Dy ions and quasi collinear magnetism of Gd in \DTG . 
The ground state energy difference between the ferromagnetic and the antiferromagnetic Dy configuration decreases relative to \DTS.   
From temperature dependent magnetization decay measurements we however infer a higher kinetic barrier.
The decrease of the prefactor for the description of the magnetic lifetime overcompensates the higher barrier and identifies Gd as a single atom catalyst for the decay of the magnetization in single molecule magnets. 

We thank A. Seitsonen for the artwork in Figure~\ref{F1} and acknowledge financial support from the Swiss National Science Foundation (SNF projects 200021L\_147201, 206021\_150784, 200020\_153312, 200021 129861, 147143, and PZ00P2-142474), the Deutsche Forschungsgemeinschaft (DFG project PO 1602/4-1 and 1602/5-1) and the Swedish Research Council (Grant No. 2015-00455). 

\bibliography{Dy2Gd}

\begin{thebibliography}{32}%
\makeatletter
\providecommand \@ifxundefined [1]{%
 \@ifx{#1\undefined}
}%
\providecommand \@ifnum [1]{%
 \ifnum #1\expandafter \@firstoftwo
 \else \expandafter \@secondoftwo
 \fi
}%
\providecommand \@ifx [1]{%
 \ifx #1\expandafter \@firstoftwo
 \else \expandafter \@secondoftwo
 \fi
}%
\providecommand \natexlab [1]{#1}%
\providecommand \enquote  [1]{``#1''}%
\providecommand \bibnamefont  [1]{#1}%
\providecommand \bibfnamefont [1]{#1}%
\providecommand \citenamefont [1]{#1}%
\providecommand \href@noop [0]{\@secondoftwo}%
\providecommand \href [0]{\begingroup \@sanitize@url \@href}%
\providecommand \@href[1]{\@@startlink{#1}\@@href}%
\providecommand \@@href[1]{\endgroup#1\@@endlink}%
\providecommand \@sanitize@url [0]{\catcode `\\12\catcode `\$12\catcode
  `\&12\catcode `\#12\catcode `\^12\catcode `\_12\catcode `\%12\relax}%
\providecommand \@@startlink[1]{}%
\providecommand \@@endlink[0]{}%
\providecommand \url  [0]{\begingroup\@sanitize@url \@url }%
\providecommand \@url [1]{\endgroup\@href {#1}{\urlprefix }}%
\providecommand \urlprefix  [0]{URL }%
\providecommand \Eprint [0]{\href }%
\providecommand \doibase [0]{http://dx.doi.org/}%
\providecommand \selectlanguage [0]{\@gobble}%
\providecommand \bibinfo  [0]{\@secondoftwo}%
\providecommand \bibfield  [0]{\@secondoftwo}%
\providecommand \translation [1]{[#1]}%
\providecommand \BibitemOpen [0]{}%
\providecommand \bibitemStop [0]{}%
\providecommand \bibitemNoStop [0]{.\EOS\space}%
\providecommand \EOS [0]{\spacefactor3000\relax}%
\providecommand \BibitemShut  [1]{\csname bibitem#1\endcsname}%
\let\auto@bib@innerbib\@empty
\bibitem [{\citenamefont {Sessoli}\ \emph {et~al.}(1993)\citenamefont
  {Sessoli}, \citenamefont {Gatteschi}, \citenamefont {Caneschi},\ and\
  \citenamefont {Novak}}]{Sessoli1993}%
  \BibitemOpen
  \bibfield  {author} {\bibinfo {author} {\bibfnamefont {R.}~\bibnamefont
  {Sessoli}}, \bibinfo {author} {\bibfnamefont {D.}~\bibnamefont {Gatteschi}},
  \bibinfo {author} {\bibfnamefont {A.}~\bibnamefont {Caneschi}}, \ and\
  \bibinfo {author} {\bibfnamefont {M.~A.}\ \bibnamefont {Novak}},\ }\bibfield
  {title} {\enquote {\bibinfo {title} {{Magnetic bistability in a metal-ion
  cluster}},}\ }\href {\doibase 10.1038/365141a0} {\bibfield  {journal}
  {\bibinfo  {journal} {Nature}\ }\textbf {\bibinfo {volume} {365}},\ \bibinfo
  {pages} {141--143} (\bibinfo {year} {1993})}\BibitemShut {NoStop}%
\bibitem [{\citenamefont {Gatteschi}\ \emph {et~al.}(2006)\citenamefont
  {Gatteschi}, \citenamefont {Sessoli},\ and\ \citenamefont
  {Villain}}]{Gatteschi2006}%
  \BibitemOpen
  \bibfield  {author} {\bibinfo {author} {\bibfnamefont {Dante}\ \bibnamefont
  {Gatteschi}}, \bibinfo {author} {\bibfnamefont {Roberta}\ \bibnamefont
  {Sessoli}}, \ and\ \bibinfo {author} {\bibfnamefont {Jacques}\ \bibnamefont
  {Villain}},\ }\href {\doibase 10.1093/acprof:oso/9780198567530.001.0001}
  {\emph {\bibinfo {title} {{Molecular Nanomagnets}}}}\ (\bibinfo  {publisher}
  {Oxford University Press},\ \bibinfo {year} {2006})\BibitemShut {NoStop}%
\bibitem [{\citenamefont {Coronado}(2019)}]{cor2019}%
  \BibitemOpen
  \bibfield  {author} {\bibinfo {author} {\bibfnamefont {Eugenio}\ \bibnamefont
  {Coronado}},\ }\bibfield  {title} {\enquote {\bibinfo {title} {Molecular
  magnetism: from chemical design to spin control in molecules, materials and
  devices},}\ }\href {\doibase 10.1038/s41578-019-0146-8} {\bibfield  {journal}
  {\bibinfo  {journal} {Nature Reviews Materials}\ } (\bibinfo {year} {2019}),\
  10.1038/s41578-019-0146-8}\BibitemShut {NoStop}%
\bibitem [{\citenamefont {Ishikawa}\ \emph {et~al.}({2003})\citenamefont
  {Ishikawa}, \citenamefont {Sugita}, \citenamefont {Ishikawa}, \citenamefont
  {Koshihara},\ and\ \citenamefont {Kaizu}}]{ish2003}%
  \BibitemOpen
  \bibfield  {author} {\bibinfo {author} {\bibfnamefont {N}~\bibnamefont
  {Ishikawa}}, \bibinfo {author} {\bibfnamefont {M}~\bibnamefont {Sugita}},
  \bibinfo {author} {\bibfnamefont {T}~\bibnamefont {Ishikawa}}, \bibinfo
  {author} {\bibfnamefont {S}~\bibnamefont {Koshihara}}, \ and\ \bibinfo
  {author} {\bibfnamefont {Y}~\bibnamefont {Kaizu}},\ }\bibfield  {title}
  {\enquote {\bibinfo {title} {{Lanthanide double-decker complexes functioning
  as magnets at the single-molecular level}},}\ }\href {\doibase
  {10.1021/ja029629n}} {\bibfield  {journal} {\bibinfo  {journal} {{J. Am.
  Chem. Soc.}}\ }\textbf {\bibinfo {volume} {{125}}},\ \bibinfo {pages}
  {{8694--8695}} (\bibinfo {year} {{2003}})}\BibitemShut {NoStop}%
\bibitem [{\citenamefont {Woodruff}\ \emph {et~al.}(2013)\citenamefont
  {Woodruff}, \citenamefont {Winpenny},\ and\ \citenamefont
  {Layfield}}]{woo2013}%
  \BibitemOpen
  \bibfield  {author} {\bibinfo {author} {\bibfnamefont {Daniel~N.}\
  \bibnamefont {Woodruff}}, \bibinfo {author} {\bibfnamefont {Richard E.~P.}\
  \bibnamefont {Winpenny}}, \ and\ \bibinfo {author} {\bibfnamefont
  {Richard~A.}\ \bibnamefont {Layfield}},\ }\bibfield  {title} {\enquote
  {\bibinfo {title} {Lanthanide single-molecule magnets},}\ }\href@noop {}
  {\bibfield  {journal} {\bibinfo  {journal} {Chemical Reviews}\ }\textbf
  {\bibinfo {volume} {113}},\ \bibinfo {pages} {5110--5148} (\bibinfo {year}
  {2013})}\BibitemShut {NoStop}%
\bibitem [{\citenamefont {Liddle}\ and\ \citenamefont {van
  Slageren}(2015)}]{lid2015}%
  \BibitemOpen
  \bibfield  {author} {\bibinfo {author} {\bibfnamefont {Stephen~T.}\
  \bibnamefont {Liddle}}\ and\ \bibinfo {author} {\bibfnamefont {Joris}\
  \bibnamefont {van Slageren}},\ }\bibfield  {title} {\enquote {\bibinfo
  {title} {Improving f-element single molecule magnets},}\ }\href@noop {}
  {\bibfield  {journal} {\bibinfo  {journal} {Chem. Soc. Rev.}\ }\textbf
  {\bibinfo {volume} {44}},\ \bibinfo {pages} {6655--6669} (\bibinfo {year}
  {2015})}\BibitemShut {NoStop}%
\bibitem [{\citenamefont {Liu}\ \emph {et~al.}(2018)\citenamefont {Liu},
  \citenamefont {Chen},\ and\ \citenamefont {Tong}}]{liu2018}%
  \BibitemOpen
  \bibfield  {author} {\bibinfo {author} {\bibfnamefont {Jun-Liang}\
  \bibnamefont {Liu}}, \bibinfo {author} {\bibfnamefont {Yan-Cong}\
  \bibnamefont {Chen}}, \ and\ \bibinfo {author} {\bibfnamefont {Ming-Liang}\
  \bibnamefont {Tong}},\ }\bibfield  {title} {\enquote {\bibinfo {title}
  {Symmetry strategies for high performance lanthanide-based single-molecule
  magnets},}\ }\href {\doibase 10.1039/C7CS00266A} {\bibfield  {journal}
  {\bibinfo  {journal} {Chem. Soc. Rev.}\ }\textbf {\bibinfo {volume} {47}},\
  \bibinfo {pages} {2431--2453} (\bibinfo {year} {2018})}\BibitemShut {NoStop}%
\bibitem [{\citenamefont {Goodwin}\ \emph {et~al.}({2017})\citenamefont
  {Goodwin}, \citenamefont {Ortu}, \citenamefont {Reta}, \citenamefont
  {Chilton},\ and\ \citenamefont {Mills}}]{goo2017}%
  \BibitemOpen
  \bibfield  {author} {\bibinfo {author} {\bibfnamefont {Conrad A.~P.}\
  \bibnamefont {Goodwin}}, \bibinfo {author} {\bibfnamefont {Fabrizio}\
  \bibnamefont {Ortu}}, \bibinfo {author} {\bibfnamefont {Daniel}\ \bibnamefont
  {Reta}}, \bibinfo {author} {\bibfnamefont {Nicholas~F.}\ \bibnamefont
  {Chilton}}, \ and\ \bibinfo {author} {\bibfnamefont {David~P.}\ \bibnamefont
  {Mills}},\ }\bibfield  {title} {\enquote {\bibinfo {title} {{Molecular
  magnetic hysteresis at 60 kelvin in dysprosocenium}},}\ }\href {\doibase
  {10.1038/nature23447}} {\bibfield  {journal} {\bibinfo  {journal} {{Nature}}\
  }\textbf {\bibinfo {volume} {{548}}},\ \bibinfo {pages} {{439--442}}
  (\bibinfo {year} {{2017}})}\BibitemShut {NoStop}%
\bibitem [{\citenamefont {Guo}\ \emph {et~al.}({2018})\citenamefont {Guo},
  \citenamefont {Day}, \citenamefont {Chen}, \citenamefont {Tong},
  \citenamefont {Mansikkamaki},\ and\ \citenamefont {Layfield}}]{guo2018}%
  \BibitemOpen
  \bibfield  {author} {\bibinfo {author} {\bibfnamefont {Fu-Sheng}\
  \bibnamefont {Guo}}, \bibinfo {author} {\bibfnamefont {Benjamin~M.}\
  \bibnamefont {Day}}, \bibinfo {author} {\bibfnamefont {Yan-Cong}\
  \bibnamefont {Chen}}, \bibinfo {author} {\bibfnamefont {Ming-Liang}\
  \bibnamefont {Tong}}, \bibinfo {author} {\bibfnamefont {Akseli}\ \bibnamefont
  {Mansikkamaki}}, \ and\ \bibinfo {author} {\bibfnamefont {Richard~A.}\
  \bibnamefont {Layfield}},\ }\bibfield  {title} {\enquote {\bibinfo {title}
  {{Magnetic hysteresis up to 80 kelvin in a dysprosium metallocene
  single-molecule magnet}},}\ }\href {\doibase {10.1126/science.aav0652}}
  {\bibfield  {journal} {\bibinfo  {journal} {{Science}}\ }\textbf {\bibinfo
  {volume} {{362}}},\ \bibinfo {pages} {{1400+}} (\bibinfo {year}
  {{2018}})}\BibitemShut {NoStop}%
\bibitem [{\citenamefont {Stevenson}\ \emph {et~al.}({1999})\citenamefont
  {Stevenson}, \citenamefont {Rice}, \citenamefont {Glass}, \citenamefont
  {Harich}, \citenamefont {Cromer}, \citenamefont {Jordan}, \citenamefont
  {Craft}, \citenamefont {Hadju}, \citenamefont {Bible}, \citenamefont
  {Olmstead}, \citenamefont {Maitra}, \citenamefont {Fisher}, \citenamefont
  {Balch},\ and\ \citenamefont {Dorn}}]{ste99}%
  \BibitemOpen
  \bibfield  {author} {\bibinfo {author} {\bibfnamefont {S}~\bibnamefont
  {Stevenson}}, \bibinfo {author} {\bibfnamefont {G}~\bibnamefont {Rice}},
  \bibinfo {author} {\bibfnamefont {T}~\bibnamefont {Glass}}, \bibinfo {author}
  {\bibfnamefont {K}~\bibnamefont {Harich}}, \bibinfo {author} {\bibfnamefont
  {F}~\bibnamefont {Cromer}}, \bibinfo {author} {\bibfnamefont
  {MR}~\bibnamefont {Jordan}}, \bibinfo {author} {\bibfnamefont
  {J}~\bibnamefont {Craft}}, \bibinfo {author} {\bibfnamefont {E}~\bibnamefont
  {Hadju}}, \bibinfo {author} {\bibfnamefont {R}~\bibnamefont {Bible}},
  \bibinfo {author} {\bibfnamefont {MM}~\bibnamefont {Olmstead}}, \bibinfo
  {author} {\bibfnamefont {K}~\bibnamefont {Maitra}}, \bibinfo {author}
  {\bibfnamefont {AJ}~\bibnamefont {Fisher}}, \bibinfo {author} {\bibfnamefont
  {AL}~\bibnamefont {Balch}}, \ and\ \bibinfo {author} {\bibfnamefont
  {HC}~\bibnamefont {Dorn}},\ }\bibfield  {title} {\enquote {\bibinfo {title}
  {{Small-bandgap endohedral metallofullerenes in high yield and purity}},}\
  }\href@noop {} {\bibfield  {journal} {\bibinfo  {journal} {{NATURE}}\
  }\textbf {\bibinfo {volume} {{401}}},\ \bibinfo {pages} {{55--57}} (\bibinfo
  {year} {{1999}})}\BibitemShut {NoStop}%
\bibitem [{\citenamefont {Popov}\ \emph {et~al.}(2013)\citenamefont {Popov},
  \citenamefont {Yang},\ and\ \citenamefont {Dunsch}}]{Popov2013}%
  \BibitemOpen
  \bibfield  {author} {\bibinfo {author} {\bibfnamefont {Alexey~A.}\
  \bibnamefont {Popov}}, \bibinfo {author} {\bibfnamefont {Shangfeng}\
  \bibnamefont {Yang}}, \ and\ \bibinfo {author} {\bibfnamefont {Lothar}\
  \bibnamefont {Dunsch}},\ }\bibfield  {title} {\enquote {\bibinfo {title}
  {{Endohedral Fullerenes}},}\ }\href {\doibase 10.1021/cr300297r} {\bibfield
  {journal} {\bibinfo  {journal} {Chem. Rev.}\ }\textbf {\bibinfo {volume}
  {113}},\ \bibinfo {pages} {5989--6113} (\bibinfo {year} {2013})}\BibitemShut
  {NoStop}%
\bibitem [{\citenamefont {Funasaka}\ \emph {et~al.}(1995)\citenamefont
  {Funasaka}, \citenamefont {Sakurai}, \citenamefont {Oda}, \citenamefont
  {Yamamoto},\ and\ \citenamefont {Takahashi}}]{fun1995}%
  \BibitemOpen
  \bibfield  {author} {\bibinfo {author} {\bibfnamefont {Hideyuki}\
  \bibnamefont {Funasaka}}, \bibinfo {author} {\bibfnamefont {Koji}\
  \bibnamefont {Sakurai}}, \bibinfo {author} {\bibfnamefont {Yoshihiro}\
  \bibnamefont {Oda}}, \bibinfo {author} {\bibfnamefont {Kazunori}\
  \bibnamefont {Yamamoto}}, \ and\ \bibinfo {author} {\bibfnamefont {Takeshi}\
  \bibnamefont {Takahashi}},\ }\bibfield  {title} {\enquote {\bibinfo {title}
  {{Magnetic properties of Gd@C$_{82}$ metallofullerene}},}\ }\href {\doibase
  https://doi.org/10.1016/0009-2614(95)90631-2} {\bibfield  {journal} {\bibinfo
   {journal} {Chemical Physics Letters}\ }\textbf {\bibinfo {volume} {232}},\
  \bibinfo {pages} {273 -- 277} (\bibinfo {year} {1995})}\BibitemShut {NoStop}%
\bibitem [{\citenamefont {Westerstr{\"{o}}m}\ \emph {et~al.}(2012)\citenamefont
  {Westerstr{\"{o}}m}, \citenamefont {Dreiser}, \citenamefont {Piamonteze},
  \citenamefont {Muntwiler}, \citenamefont {Weyeneth}, \citenamefont {Brune},
  \citenamefont {Rusponi}, \citenamefont {Nolting}, \citenamefont {Popov},
  \citenamefont {Yang}, \citenamefont {Dunsch},\ and\ \citenamefont
  {Greber}}]{Westerstrom2012}%
  \BibitemOpen
  \bibfield  {author} {\bibinfo {author} {\bibfnamefont {Rasmus}\ \bibnamefont
  {Westerstr{\"{o}}m}}, \bibinfo {author} {\bibfnamefont {Jan}\ \bibnamefont
  {Dreiser}}, \bibinfo {author} {\bibfnamefont {Cinthia}\ \bibnamefont
  {Piamonteze}}, \bibinfo {author} {\bibfnamefont {Matthias}\ \bibnamefont
  {Muntwiler}}, \bibinfo {author} {\bibfnamefont {Stephen}\ \bibnamefont
  {Weyeneth}}, \bibinfo {author} {\bibfnamefont {Harald}\ \bibnamefont
  {Brune}}, \bibinfo {author} {\bibfnamefont {Stefano}\ \bibnamefont
  {Rusponi}}, \bibinfo {author} {\bibfnamefont {Frithjof}\ \bibnamefont
  {Nolting}}, \bibinfo {author} {\bibfnamefont {Alexey}\ \bibnamefont {Popov}},
  \bibinfo {author} {\bibfnamefont {Shangfeng}\ \bibnamefont {Yang}}, \bibinfo
  {author} {\bibfnamefont {Lothar}\ \bibnamefont {Dunsch}}, \ and\ \bibinfo
  {author} {\bibfnamefont {Thomas}\ \bibnamefont {Greber}},\ }\bibfield
  {title} {\enquote {\bibinfo {title} {{An Endohedral Single-Molecule Magnet
  with Long Relaxation Times: DySc\(_{2}\)N@C\(_{80}\)}},}\ }\href {\doibase
  10.1021/ja301044p} {\bibfield  {journal} {\bibinfo  {journal} {J. Am. Chem.
  Soc.}\ }\textbf {\bibinfo {volume} {134}},\ \bibinfo {pages} {9840--9843}
  (\bibinfo {year} {2012})}\BibitemShut {NoStop}%
\bibitem [{\citenamefont {Junghans}\ \emph {et~al.}({2015})\citenamefont
  {Junghans}, \citenamefont {Schlesier}, \citenamefont {Kostanyan},
  \citenamefont {Samoylova}, \citenamefont {Deng}, \citenamefont {Rosenkranz},
  \citenamefont {Schiemenz}, \citenamefont {Westerstroem}, \citenamefont
  {Greber}, \citenamefont {Buechner},\ and\ \citenamefont
  {Popov}}]{Junghans2015}%
  \BibitemOpen
  \bibfield  {author} {\bibinfo {author} {\bibfnamefont {Katrin}\ \bibnamefont
  {Junghans}}, \bibinfo {author} {\bibfnamefont {Christin}\ \bibnamefont
  {Schlesier}}, \bibinfo {author} {\bibfnamefont {Aram}\ \bibnamefont
  {Kostanyan}}, \bibinfo {author} {\bibfnamefont {Nataliya~A.}\ \bibnamefont
  {Samoylova}}, \bibinfo {author} {\bibfnamefont {Qingming}\ \bibnamefont
  {Deng}}, \bibinfo {author} {\bibfnamefont {Marco}\ \bibnamefont
  {Rosenkranz}}, \bibinfo {author} {\bibfnamefont {Sandra}\ \bibnamefont
  {Schiemenz}}, \bibinfo {author} {\bibfnamefont {Rasmus}\ \bibnamefont
  {Westerstroem}}, \bibinfo {author} {\bibfnamefont {Thomas}\ \bibnamefont
  {Greber}}, \bibinfo {author} {\bibfnamefont {Bernd}\ \bibnamefont
  {Buechner}}, \ and\ \bibinfo {author} {\bibfnamefont {Alexey~A.}\
  \bibnamefont {Popov}},\ }\bibfield  {title} {\enquote {\bibinfo {title}
  {{Methane as a Selectivity Booster in the Arc-Discharge Synthesis of
  Endohedral Fullerenes: Selective Synthesis of the Single-Molecule Magnet
  Dy$_2$TiC@C$_{80}$ and its Congener Dy$_2$TiC$_2$@C$_{80}$}},}\ }\href
  {\doibase {10.1002/anie.201505870}} {\bibfield  {journal} {\bibinfo
  {journal} {{Angewandte Chemie International Edition}}\ }\textbf {\bibinfo
  {volume} {{54}}},\ \bibinfo {pages} {{13411--13415}} (\bibinfo {year}
  {{2015}})}\BibitemShut {NoStop}%
\bibitem [{\citenamefont {Schlesier}\ \emph {et~al.}({2019})\citenamefont
  {Schlesier}, \citenamefont {Liu}, \citenamefont {Dubrovin}, \citenamefont
  {Spree}, \citenamefont {Buechner}, \citenamefont {Avdoshenko},\ and\
  \citenamefont {Popov}}]{Schlesier2019}%
  \BibitemOpen
  \bibfield  {author} {\bibinfo {author} {\bibfnamefont {C.}~\bibnamefont
  {Schlesier}}, \bibinfo {author} {\bibfnamefont {F.}~\bibnamefont {Liu}},
  \bibinfo {author} {\bibfnamefont {V.}~\bibnamefont {Dubrovin}}, \bibinfo
  {author} {\bibfnamefont {L.}~\bibnamefont {Spree}}, \bibinfo {author}
  {\bibfnamefont {B.}~\bibnamefont {Buechner}}, \bibinfo {author}
  {\bibfnamefont {S.~M.}\ \bibnamefont {Avdoshenko}}, \ and\ \bibinfo {author}
  {\bibfnamefont {A.~A.}\ \bibnamefont {Popov}},\ }\bibfield  {title} {\enquote
  {\bibinfo {title} {{Mixed dysprosium-lanthanide nitride clusterfullerenes
  DyM$_2$N@C$_{80}$ I$_h$ and Dy$_2$MN@C$_{80}$ I$_h$ (M = Gd, Er, Tm, and Lu):
  synthesis, molecular structure, and quantum motion of the endohedral nitrogen
  atom}},}\ }\href {\doibase {10.1039/c9nr03593a}} {\bibfield  {journal}
  {\bibinfo  {journal} {{Nanoscale}}\ }\textbf {\bibinfo {volume} {{11}}},\
  \bibinfo {pages} {{13139--13153}} (\bibinfo {year} {{2019}})}\BibitemShut
  {NoStop}%
\bibitem [{\citenamefont {Spree}\ \emph {et~al.}({2020})\citenamefont {Spree},
  \citenamefont {Schlesier}, \citenamefont {Kostanyan}, \citenamefont
  {Westerstroem}, \citenamefont {Greber}, \citenamefont {Buechner},
  \citenamefont {Avdoshenko},\ and\ \citenamefont {Popov}}]{Spree2020}%
  \BibitemOpen
  \bibfield  {author} {\bibinfo {author} {\bibfnamefont {Lukas}\ \bibnamefont
  {Spree}}, \bibinfo {author} {\bibfnamefont {Christin}\ \bibnamefont
  {Schlesier}}, \bibinfo {author} {\bibfnamefont {Aram}\ \bibnamefont
  {Kostanyan}}, \bibinfo {author} {\bibfnamefont {Rasmus}\ \bibnamefont
  {Westerstroem}}, \bibinfo {author} {\bibfnamefont {Thomas}\ \bibnamefont
  {Greber}}, \bibinfo {author} {\bibfnamefont {Bernd}\ \bibnamefont
  {Buechner}}, \bibinfo {author} {\bibfnamefont {Stanislav~M.}\ \bibnamefont
  {Avdoshenko}}, \ and\ \bibinfo {author} {\bibfnamefont {Alexey~A.}\
  \bibnamefont {Popov}},\ }\bibfield  {title} {\enquote {\bibinfo {title}
  {{Single-Molecule Magnets DyM$_2$N@C$_{80}$ and Dy$_2$MN@$_{80}$ (M=Sc, Lu):
  The Impact of Diamagnetic Metals on Dy$^{3+}$ Magnetic Anisotropy, Dy center
  dot center dot Dy Coupling, and Mixing of Molecular and Lattice
  Vibrations}},}\ }\href {\doibase {10.1002/chem.201904879}} {\bibfield
  {journal} {\bibinfo  {journal} {{Chemistry-A European Journal}}\ }\textbf
  {\bibinfo {volume} {{26}}},\ \bibinfo {pages} {{2436--2449}} (\bibinfo {year}
  {{2020}})}\BibitemShut {NoStop}%
\bibitem [{\citenamefont {Rinehart}\ \emph {et~al.}({2011})\citenamefont
  {Rinehart}, \citenamefont {Fang}, \citenamefont {Evans},\ and\ \citenamefont
  {Long}}]{rin2011}%
  \BibitemOpen
  \bibfield  {author} {\bibinfo {author} {\bibfnamefont {Jeffrey~D.}\
  \bibnamefont {Rinehart}}, \bibinfo {author} {\bibfnamefont {Ming}\
  \bibnamefont {Fang}}, \bibinfo {author} {\bibfnamefont {William~J.}\
  \bibnamefont {Evans}}, \ and\ \bibinfo {author} {\bibfnamefont {Jeffrey~R.}\
  \bibnamefont {Long}},\ }\bibfield  {title} {\enquote {\bibinfo {title}
  {{Strong exchange and magnetic blocking in N$_2^{3-}$-radical-bridged
  lanthanide complexes}},}\ }\href {\doibase {10.1038/nchem.1063}} {\bibfield
  {journal} {\bibinfo  {journal} {{Nature Chem.}}\ }\textbf {\bibinfo {volume}
  {{3}}},\ \bibinfo {pages} {{538--542}} (\bibinfo {year}
  {{2011}})}\BibitemShut {NoStop}%
\bibitem [{\citenamefont {Spree}\ and\ \citenamefont {Popov}(2019)}]{spr2019}%
  \BibitemOpen
  \bibfield  {author} {\bibinfo {author} {\bibfnamefont {Lukas}\ \bibnamefont
  {Spree}}\ and\ \bibinfo {author} {\bibfnamefont {Alexey~A.}\ \bibnamefont
  {Popov}},\ }\bibfield  {title} {\enquote {\bibinfo {title} {Recent advances
  in single molecule magnetism of dysprosium-metallofullerenes},}\ }\href@noop
  {} {\bibfield  {journal} {\bibinfo  {journal} {Dalton Trans.}\ }\textbf
  {\bibinfo {volume} {48}},\ \bibinfo {pages} {2861--2871} (\bibinfo {year}
  {2019})}\BibitemShut {NoStop}%
\bibitem [{\citenamefont {Westerstr{\"{o}}m}\ \emph {et~al.}(2014)\citenamefont
  {Westerstr{\"{o}}m}, \citenamefont {Dreiser}, \citenamefont {Piamonteze},
  \citenamefont {Muntwiler}, \citenamefont {Weyeneth}, \citenamefont
  {Kr{\"{a}}mer}, \citenamefont {Liu}, \citenamefont {Decurtins}, \citenamefont
  {Popov}, \citenamefont {Yang}, \citenamefont {Dunsch},\ and\ \citenamefont
  {Greber}}]{Westerstrom2014}%
  \BibitemOpen
  \bibfield  {author} {\bibinfo {author} {\bibfnamefont {Rasmus}\ \bibnamefont
  {Westerstr{\"{o}}m}}, \bibinfo {author} {\bibfnamefont {Jan}\ \bibnamefont
  {Dreiser}}, \bibinfo {author} {\bibfnamefont {Cinthia}\ \bibnamefont
  {Piamonteze}}, \bibinfo {author} {\bibfnamefont {Matthias}\ \bibnamefont
  {Muntwiler}}, \bibinfo {author} {\bibfnamefont {Stephen}\ \bibnamefont
  {Weyeneth}}, \bibinfo {author} {\bibfnamefont {Karl}\ \bibnamefont
  {Kr{\"{a}}mer}}, \bibinfo {author} {\bibfnamefont {Shi-Xia}\ \bibnamefont
  {Liu}}, \bibinfo {author} {\bibfnamefont {Silvio}\ \bibnamefont {Decurtins}},
  \bibinfo {author} {\bibfnamefont {Alexey}\ \bibnamefont {Popov}}, \bibinfo
  {author} {\bibfnamefont {Shangfeng}\ \bibnamefont {Yang}}, \bibinfo {author}
  {\bibfnamefont {Lothar}\ \bibnamefont {Dunsch}}, \ and\ \bibinfo {author}
  {\bibfnamefont {Thomas}\ \bibnamefont {Greber}},\ }\bibfield  {title}
  {\enquote {\bibinfo {title} {{Tunneling, remanence, and frustration in
  dysprosium-based endohedral single-molecule magnets}},}\ }\href {\doibase
  10.1103/PhysRevB.89.060406} {\bibfield  {journal} {\bibinfo  {journal} {Phys.
  Rev. B}\ }\textbf {\bibinfo {volume} {89}},\ \bibinfo {pages} {060406}
  (\bibinfo {year} {2014})}\BibitemShut {NoStop}%
\bibitem [{\citenamefont {Nie}\ \emph {et~al.}({2019})\citenamefont {Nie},
  \citenamefont {Xiong}, \citenamefont {Zhao}, \citenamefont {Meng},
  \citenamefont {Zhang}, \citenamefont {Han}, \citenamefont {Li}, \citenamefont
  {Wang}, \citenamefont {Feng}, \citenamefont {Wang},\ and\ \citenamefont
  {Wang}}]{Nie2019}%
  \BibitemOpen
  \bibfield  {author} {\bibinfo {author} {\bibfnamefont {Mingzhe}\ \bibnamefont
  {Nie}}, \bibinfo {author} {\bibfnamefont {Jin}\ \bibnamefont {Xiong}},
  \bibinfo {author} {\bibfnamefont {Chong}\ \bibnamefont {Zhao}}, \bibinfo
  {author} {\bibfnamefont {Haibing}\ \bibnamefont {Meng}}, \bibinfo {author}
  {\bibfnamefont {Kun}\ \bibnamefont {Zhang}}, \bibinfo {author} {\bibfnamefont
  {Yibo}\ \bibnamefont {Han}}, \bibinfo {author} {\bibfnamefont {Jie}\
  \bibnamefont {Li}}, \bibinfo {author} {\bibfnamefont {Bingwu}\ \bibnamefont
  {Wang}}, \bibinfo {author} {\bibfnamefont {Lai}\ \bibnamefont {Feng}},
  \bibinfo {author} {\bibfnamefont {Chunru}\ \bibnamefont {Wang}}, \ and\
  \bibinfo {author} {\bibfnamefont {Taishan}\ \bibnamefont {Wang}},\ }\bibfield
   {title} {\enquote {\bibinfo {title} {{Luminescent single-molecule magnet of
  metallofullerene DyErScN@I$_{h}$C$_{80}$}},}\ }\href {\doibase
  {10.1007/s12274-019-2429-1}} {\bibfield  {journal} {\bibinfo  {journal}
  {{Nano Research}}\ }\textbf {\bibinfo {volume} {{12}}},\ \bibinfo {pages}
  {{1727--1731}} (\bibinfo {year} {{2019}})}\BibitemShut {NoStop}%
\bibitem [{\citenamefont {van~der Laan}\ and\ \citenamefont
  {Figueroa}(2014)}]{van2014}%
  \BibitemOpen
  \bibfield  {author} {\bibinfo {author} {\bibfnamefont {Gerrit}\ \bibnamefont
  {van~der Laan}}\ and\ \bibinfo {author} {\bibfnamefont {Adriana~I.}\
  \bibnamefont {Figueroa}},\ }\bibfield  {title} {\enquote {\bibinfo {title}
  {X-ray magnetic circular dichroismÑa versatile tool to study magnetism},}\
  }\href {\doibase https://doi.org/10.1016/j.ccr.2014.03.018} {\bibfield
  {journal} {\bibinfo  {journal} {Coordination Chemistry Reviews}\ }\textbf
  {\bibinfo {volume} {277-278}},\ \bibinfo {pages} {95 -- 129} (\bibinfo {year}
  {2014})}\BibitemShut {NoStop}%
\bibitem [{\citenamefont {Dreiser}\ \emph {et~al.}(2012)\citenamefont
  {Dreiser}, \citenamefont {Pedersen}, \citenamefont {Piamonteze},
  \citenamefont {Rusponi}, \citenamefont {Salman}, \citenamefont {Ali},
  \citenamefont {Schau-Magnussen}, \citenamefont {Thuesen}, \citenamefont
  {Piligkos}, \citenamefont {Weihe}, \citenamefont {Mutka}, \citenamefont
  {Waldmann}, \citenamefont {Oppeneer}, \citenamefont {Bendix}, \citenamefont
  {Nolting},\ and\ \citenamefont {Brune}}]{dre2012}%
  \BibitemOpen
  \bibfield  {author} {\bibinfo {author} {\bibfnamefont {Jan}\ \bibnamefont
  {Dreiser}}, \bibinfo {author} {\bibfnamefont {Kasper~S.}\ \bibnamefont
  {Pedersen}}, \bibinfo {author} {\bibfnamefont {Cinthia}\ \bibnamefont
  {Piamonteze}}, \bibinfo {author} {\bibfnamefont {Stefano}\ \bibnamefont
  {Rusponi}}, \bibinfo {author} {\bibfnamefont {Zaher}\ \bibnamefont {Salman}},
  \bibinfo {author} {\bibfnamefont {Md.~Ehesan}\ \bibnamefont {Ali}}, \bibinfo
  {author} {\bibfnamefont {Magnus}\ \bibnamefont {Schau-Magnussen}}, \bibinfo
  {author} {\bibfnamefont {Christian~Aa.}\ \bibnamefont {Thuesen}}, \bibinfo
  {author} {\bibfnamefont {Stergios}\ \bibnamefont {Piligkos}}, \bibinfo
  {author} {\bibfnamefont {H¿gni}\ \bibnamefont {Weihe}}, \bibinfo {author}
  {\bibfnamefont {Hannu}\ \bibnamefont {Mutka}}, \bibinfo {author}
  {\bibfnamefont {Oliver}\ \bibnamefont {Waldmann}}, \bibinfo {author}
  {\bibfnamefont {Peter}\ \bibnamefont {Oppeneer}}, \bibinfo {author}
  {\bibfnamefont {Jesper}\ \bibnamefont {Bendix}}, \bibinfo {author}
  {\bibfnamefont {Frithjof}\ \bibnamefont {Nolting}}, \ and\ \bibinfo {author}
  {\bibfnamefont {Harald}\ \bibnamefont {Brune}},\ }\bibfield  {title}
  {\enquote {\bibinfo {title} {Direct observation of a ferri-to-ferromagnetic
  transition in a fluoride-bridged 3dÐ4f molecular cluster},}\ }\href {\doibase
  10.1039/C2SC00794K} {\bibfield  {journal} {\bibinfo  {journal} {Chem. Sci.}\
  }\textbf {\bibinfo {volume} {3}},\ \bibinfo {pages} {1024--1032} (\bibinfo
  {year} {2012})}\BibitemShut {NoStop}%
\bibitem [{\citenamefont {Corradini}\ \emph {et~al.}(2012)\citenamefont
  {Corradini}, \citenamefont {Ghirri}, \citenamefont {Garlatti}, \citenamefont
  {Biagi}, \citenamefont {De~Renzi}, \citenamefont {del Pennino}, \citenamefont
  {Bellini}, \citenamefont {Carretta}, \citenamefont {Santini}, \citenamefont
  {Timco}, \citenamefont {Winpenny},\ and\ \citenamefont {Affronte}}]{cor2012}%
  \BibitemOpen
  \bibfield  {author} {\bibinfo {author} {\bibfnamefont {Valdis}\ \bibnamefont
  {Corradini}}, \bibinfo {author} {\bibfnamefont {Alberto}\ \bibnamefont
  {Ghirri}}, \bibinfo {author} {\bibfnamefont {Elena}\ \bibnamefont
  {Garlatti}}, \bibinfo {author} {\bibfnamefont {Roberto}\ \bibnamefont
  {Biagi}}, \bibinfo {author} {\bibfnamefont {Valentina}\ \bibnamefont
  {De~Renzi}}, \bibinfo {author} {\bibfnamefont {Umberto}\ \bibnamefont {del
  Pennino}}, \bibinfo {author} {\bibfnamefont {Valerio}\ \bibnamefont
  {Bellini}}, \bibinfo {author} {\bibfnamefont {Stefano}\ \bibnamefont
  {Carretta}}, \bibinfo {author} {\bibfnamefont {Paolo}\ \bibnamefont
  {Santini}}, \bibinfo {author} {\bibfnamefont {Grigore}\ \bibnamefont
  {Timco}}, \bibinfo {author} {\bibfnamefont {Richard E.~P.}\ \bibnamefont
  {Winpenny}}, \ and\ \bibinfo {author} {\bibfnamefont {Marco}\ \bibnamefont
  {Affronte}},\ }\bibfield  {title} {\enquote {\bibinfo {title} {{Magnetic
  Anisotropy of Cr$_7$Ni Spin Clusters on Surfaces}},}\ }\href {\doibase
  10.1002/adfm.201200478} {\bibfield  {journal} {\bibinfo  {journal} {Advanced
  Functional Materials}\ }\textbf {\bibinfo {volume} {22}},\ \bibinfo {pages}
  {3706--3713} (\bibinfo {year} {2012})}\BibitemShut {NoStop}%
\bibitem [{luz()}]{luz2013}%
  \BibitemOpen
  \href@noop {} {\ }\BibitemShut {NoStop}%
\bibitem [{sup()}]{supplementals}%
  \BibitemOpen
  \href@noop {} {}\bibinfo {note} {See Supplementary Material}\BibitemShut
  {NoStop}%
\bibitem [{\citenamefont {Piamonteze}\ \emph {et~al.}(2012)\citenamefont
  {Piamonteze}, \citenamefont {Flechsig}, \citenamefont {Rusponi},
  \citenamefont {Dreiser}, \citenamefont {Heidler}, \citenamefont {Schmidt},
  \citenamefont {Wetter}, \citenamefont {Calvi}, \citenamefont {Schmidt},
  \citenamefont {Pruchova}, \citenamefont {Krempasky}, \citenamefont
  {Quitmann}, \citenamefont {Brune},\ and\ \citenamefont
  {Nolting}}]{Piamonteze2012}%
  \BibitemOpen
  \bibfield  {author} {\bibinfo {author} {\bibfnamefont {Cinthia}\ \bibnamefont
  {Piamonteze}}, \bibinfo {author} {\bibfnamefont {Uwe}\ \bibnamefont
  {Flechsig}}, \bibinfo {author} {\bibfnamefont {Stefano}\ \bibnamefont
  {Rusponi}}, \bibinfo {author} {\bibfnamefont {Jan}\ \bibnamefont {Dreiser}},
  \bibinfo {author} {\bibfnamefont {Jakoba}\ \bibnamefont {Heidler}}, \bibinfo
  {author} {\bibfnamefont {Marcus}\ \bibnamefont {Schmidt}}, \bibinfo {author}
  {\bibfnamefont {Reto}\ \bibnamefont {Wetter}}, \bibinfo {author}
  {\bibfnamefont {Marco}\ \bibnamefont {Calvi}}, \bibinfo {author}
  {\bibfnamefont {Thomas}\ \bibnamefont {Schmidt}}, \bibinfo {author}
  {\bibfnamefont {Helena}\ \bibnamefont {Pruchova}}, \bibinfo {author}
  {\bibfnamefont {Juraj}\ \bibnamefont {Krempasky}}, \bibinfo {author}
  {\bibfnamefont {Christoph}\ \bibnamefont {Quitmann}}, \bibinfo {author}
  {\bibfnamefont {Harald}\ \bibnamefont {Brune}}, \ and\ \bibinfo {author}
  {\bibfnamefont {Frithjof}\ \bibnamefont {Nolting}},\ }\bibfield  {title}
  {\enquote {\bibinfo {title} {{X-Treme beamline at SLS: X-ray magnetic
  circular and linear dichroism at high field and low temperature}},}\ }\href
  {\doibase 10.1107/S0909049512027847} {\bibfield  {journal} {\bibinfo
  {journal} {J. Synchrotron Radiat.}\ }\textbf {\bibinfo {volume} {19}},\
  \bibinfo {pages} {661--674} (\bibinfo {year} {2012})}\BibitemShut {NoStop}%
\bibitem [{\citenamefont {Jiang}\ \emph {et~al.}(2011)\citenamefont {Jiang},
  \citenamefont {Wang}, \citenamefont {Sun}, \citenamefont {Wang},\ and\
  \citenamefont {Gao}}]{Jiang2011}%
  \BibitemOpen
  \bibfield  {author} {\bibinfo {author} {\bibfnamefont {Shang-Da}\
  \bibnamefont {Jiang}}, \bibinfo {author} {\bibfnamefont {Bing-Wu}\
  \bibnamefont {Wang}}, \bibinfo {author} {\bibfnamefont {Hao-Ling}\
  \bibnamefont {Sun}}, \bibinfo {author} {\bibfnamefont {Zhe-Ming}\
  \bibnamefont {Wang}}, \ and\ \bibinfo {author} {\bibfnamefont {Song}\
  \bibnamefont {Gao}},\ }\bibfield  {title} {\enquote {\bibinfo {title} {{An
  Organometallic Single-Ion Magnet}},}\ }\href {\doibase 10.1021/ja200198v}
  {\bibfield  {journal} {\bibinfo  {journal} {J. Am. Chem. Soc.}\ }\textbf
  {\bibinfo {volume} {133}},\ \bibinfo {pages} {4730--4733} (\bibinfo {year}
  {2011})}\BibitemShut {NoStop}%
\bibitem [{\citenamefont {Teramura}\ \emph {et~al.}(1996)\citenamefont
  {Teramura}, \citenamefont {Tanaka}, \citenamefont {Thole},\ and\
  \citenamefont {Jo}}]{Teramura1996}%
  \BibitemOpen
  \bibfield  {author} {\bibinfo {author} {\bibfnamefont {Yoshiki}\ \bibnamefont
  {Teramura}}, \bibinfo {author} {\bibfnamefont {Arata}\ \bibnamefont
  {Tanaka}}, \bibinfo {author} {\bibfnamefont {B.~T.}\ \bibnamefont {Thole}}, \
  and\ \bibinfo {author} {\bibfnamefont {Takeo}\ \bibnamefont {Jo}},\
  }\bibfield  {title} {\enquote {\bibinfo {title} {{Effect of Coulomb
  Interaction on the X-Ray Magnetic Circular Dichroism Spin Sum Rule in Rare
  Earths}},}\ }\href {\doibase 10.1143/JPSJ.65.3056} {\bibfield  {journal}
  {\bibinfo  {journal} {J. Phys. Soc. Japan}\ }\textbf {\bibinfo {volume}
  {65}},\ \bibinfo {pages} {3056--3059} (\bibinfo {year} {1996})}\BibitemShut
  {NoStop}%
\bibitem [{\citenamefont {Thole}\ \emph {et~al.}(1992)\citenamefont {Thole},
  \citenamefont {Carra}, \citenamefont {Sette},\ and\ \citenamefont {van~der
  Laan}}]{Thole1992}%
  \BibitemOpen
  \bibfield  {author} {\bibinfo {author} {\bibfnamefont {B.~T.}\ \bibnamefont
  {Thole}}, \bibinfo {author} {\bibfnamefont {P.}~\bibnamefont {Carra}},
  \bibinfo {author} {\bibfnamefont {F.}~\bibnamefont {Sette}}, \ and\ \bibinfo
  {author} {\bibfnamefont {G.}~\bibnamefont {van~der Laan}},\ }\bibfield
  {title} {\enquote {\bibinfo {title} {{X-ray circular dichroism as a probe of
  orbital magnetization}},}\ }\href {\doibase 10.1103/PhysRevLett.68.1943}
  {\bibfield  {journal} {\bibinfo  {journal} {Phys. Rev. Lett.}\ }\textbf
  {\bibinfo {volume} {68}},\ \bibinfo {pages} {1943--1946} (\bibinfo {year}
  {1992})}\BibitemShut {NoStop}%
\bibitem [{\citenamefont {Carra}\ \emph {et~al.}(1993)\citenamefont {Carra},
  \citenamefont {Thole}, \citenamefont {Altarelli},\ and\ \citenamefont
  {Wang}}]{Carra1993}%
  \BibitemOpen
  \bibfield  {author} {\bibinfo {author} {\bibfnamefont {Paolo}\ \bibnamefont
  {Carra}}, \bibinfo {author} {\bibfnamefont {B.~T.}\ \bibnamefont {Thole}},
  \bibinfo {author} {\bibfnamefont {Massimo}\ \bibnamefont {Altarelli}}, \ and\
  \bibinfo {author} {\bibfnamefont {Xindong}\ \bibnamefont {Wang}},\ }\bibfield
   {title} {\enquote {\bibinfo {title} {{X-ray circular dichroism and local
  magnetic fields}},}\ }\href {\doibase 10.1103/PhysRevLett.70.694} {\bibfield
  {journal} {\bibinfo  {journal} {Phys. Rev. Lett.}\ }\textbf {\bibinfo
  {volume} {70}},\ \bibinfo {pages} {694--697} (\bibinfo {year}
  {1993})}\BibitemShut {NoStop}%
\bibitem [{\citenamefont {De~Nadai}\ \emph {et~al.}({2004})\citenamefont
  {De~Nadai}, \citenamefont {Mirone}, \citenamefont {Dhesi}, \citenamefont
  {Bencok}, \citenamefont {Brookes}, \citenamefont {Marenne}, \citenamefont
  {Rudolf}, \citenamefont {Tagmatarchis}, \citenamefont {Shinohara},\ and\
  \citenamefont {Dennis}}]{den04}%
  \BibitemOpen
  \bibfield  {author} {\bibinfo {author} {\bibfnamefont {C}~\bibnamefont
  {De~Nadai}}, \bibinfo {author} {\bibfnamefont {A}~\bibnamefont {Mirone}},
  \bibinfo {author} {\bibfnamefont {SS}~\bibnamefont {Dhesi}}, \bibinfo
  {author} {\bibfnamefont {P}~\bibnamefont {Bencok}}, \bibinfo {author}
  {\bibfnamefont {NB}~\bibnamefont {Brookes}}, \bibinfo {author} {\bibfnamefont
  {I}~\bibnamefont {Marenne}}, \bibinfo {author} {\bibfnamefont
  {P}~\bibnamefont {Rudolf}}, \bibinfo {author} {\bibfnamefont {N}~\bibnamefont
  {Tagmatarchis}}, \bibinfo {author} {\bibfnamefont {H}~\bibnamefont
  {Shinohara}}, \ and\ \bibinfo {author} {\bibfnamefont {TJS}\ \bibnamefont
  {Dennis}},\ }\bibfield  {title} {\enquote {\bibinfo {title} {{Local magnetism
  in rare-earth metals encapsulated in fullerenes}},}\ }\href {\doibase
  {10.1103/PhysRevB.69.184421}} {\bibfield  {journal} {\bibinfo  {journal}
  {{Physical Review B}}\ }\textbf {\bibinfo {volume} {{69}}} (\bibinfo {year}
  {{2004}}),\ {10.1103/PhysRevB.69.184421}}\BibitemShut {NoStop}%
\bibitem [{\citenamefont {Kostanyan}\ \emph {et~al.}({2020})\citenamefont
  {Kostanyan}, \citenamefont {Westerstrom}, \citenamefont {Kunhardt},
  \citenamefont {Buchner}, \citenamefont {Popov},\ and\ \citenamefont
  {Greber}}]{Kostanyan2020}%
  \BibitemOpen
  \bibfield  {author} {\bibinfo {author} {\bibfnamefont {Aram}\ \bibnamefont
  {Kostanyan}}, \bibinfo {author} {\bibfnamefont {Rasmus}\ \bibnamefont
  {Westerstrom}}, \bibinfo {author} {\bibfnamefont {David}\ \bibnamefont
  {Kunhardt}}, \bibinfo {author} {\bibfnamefont {Bernd}\ \bibnamefont
  {Buchner}}, \bibinfo {author} {\bibfnamefont {Alexey~A.}\ \bibnamefont
  {Popov}}, \ and\ \bibinfo {author} {\bibfnamefont {Thomas}\ \bibnamefont
  {Greber}},\ }\bibfield  {title} {\enquote {\bibinfo {title} {{Sub-Kelvin
  hysteresis of the dilanthanide single-molecule magnet Tb$_2$ScN@C$_{80}$}},}\
  }\href {\doibase {10.1103/PhysRevB.101.134429}} {\bibfield  {journal}
  {\bibinfo  {journal} {{Physical Review B}}\ }\textbf {\bibinfo {volume}
  {{101}}} (\bibinfo {year} {{2020}}),\
  {10.1103/PhysRevB.101.134429}}\BibitemShut {NoStop}%
\end{thebibliography}%

\end{document}